\documentclass[pra,aps,twocolumn,a4paper,showpacs,superscriptaddress,10pt]{revtex4-1}

\usepackage{amsmath}
\usepackage{amssymb}
\usepackage{graphicx}
\usepackage{color}
\usepackage{hyperref}
\usepackage[cm]{fullpage}

\newcommand{\mytitle}{Spin-Orbit Effects in Atomic High-Harmonic Generation}

\newcommand{\rmpdfinfo}{\special{ps:: userdict /pdfmark /cleartomark load put}}

\definecolor{MyDarkGreen}{rgb}{0,0.6,0}
\definecolor{MyDarkBlue}{rgb}{0,0,0.8}
\definecolor{MyDarkRed}{rgb}{0.6,0,0.3}
\hypersetup{breaklinks=true, colorlinks=true,plainpages=true, linktocpage=true, linkcolor=MyDarkBlue, citecolor=MyDarkGreen, urlcolor=MyDarkRed, pdfborder={0 0 0},%
pdfauthor={Stefan Pabst},%
pdfsubject={Research article},
pdftitle={\mytitle}%
}

\newcommand{\abs}[1]{\left|#1\right|}
\newcommand{\redmom}[3]{\left<#1 \,\right\| #2 \left\|\, #3 \right>}

\newcommand{\ds}[0]{\displaystyle}
\newcommand{\half}[0]{\frac{1}{2}}
\newcommand{\ket}[1]{\left|#1\right>}
\newcommand{\bra}[1]{\left<#1\right|}
\newcommand{\sket}[1]{\left|#1\right)}
\newcommand{\sbra}[1]{\left(#1\right|}
\newcommand{\braket}[2]{\left<#1|#2\right>}
\newcommand{\wigsixj}[6]{\left\{\begin{array}{ccc} #1 & #2 & #3 \\ #4 & #5 & #6 \end{array}\right\}}

\begin{document}

\title{\mytitle}

\author{Stefan Pabst$^1$ and Robin Santra$^{1,2}$}

\address{$^1$ Center for Free-Electron Laser Science, DESY, Notkestrasse 85, 22607 Hamburg, Germany}
\address{$^2$ Department of Physics,University of Hamburg, Jungiusstrasse 9, 20355 Hamburg, Germany}



\date{\today}


\begin{abstract}
Spin-orbit interactions lead to small energy gaps between the outer-most $p_{1/2}$ and $p_{3/2}$  shells of noble gas atoms.
Strong-field pulses tunnel-ionize an electron out of either shell resulting in spin-orbit-driven hole motion.
These hole dynamics affect the HHG yield.
However, the spectral shape as well as the angular distribution of the HHG emission is not influenced by spin-orbit coupling.
We demonstrate the spin-orbit effect on atomic krypton by solving the multi-electron Schr\"odinger equation with the time-dependent configuration-interaction singles (TDCIS) approach.
We also provide pulse parameters where this effect can be identified in experiments through an enhancement in the HHG yield as the wavelength of the strong-field pulse increases.
\end{abstract} 

\maketitle


\section{Introduction}
\label{s1}

High-harmonic generation (HHG) has become a promising tool to study electronic structure~\cite{ItLe-Nature432-2004,KaSa-Nature-2005,VoNe-NatPhys-2011,HaCa-JPhysB-2011} and dynamics~\cite{WiGo-Science-2011,ShSo-Nature-2012,ShTi-PRL-2012}.
Many theoretical approaches have been developed to describe HHG ranging from multi-electron descriptions~\cite{Pa-EPJST-2013,BrRo-PRA-2012,GrSa-PRA-2010} to single-active-electron models~\cite{AwDe-PRA-2008,HiFa-PRA-2011} and to semiclassical models~\cite{LeCo-PRA-1994,LaCi-PRL-2013}.
Especially the semiclassical approaches are attractive because they provide a very clear picture and make clear and direct connection to measurable quantities.
Here, one main assumption is usually made: the generated ionic hole state is confined to the outermost valence shell.
In noble gas atoms, where the outermost shell is a $p$ shell with 3 degenerate states (ignoring spin-orbit coupling), a further approximation is made that the hole is localized in the orbital aligned with the laser polarization direction~\cite{PPT-JETP-1966,Po-PhUs-2004,IvSp-JMO-2005}.

This assumption holds well for light noble gas atoms usually used for producing attosecond pulses.
For heavier atoms and molecules this simplification does not generally hold anymore because of two reasons:
First, the outer-most valence shells have very similar ionization potentials such that the strong laser field can ionize an electron from multiple valence shells~\cite{McGu-Science-2008,SmIv-Nature-2009,MaHi-PRL104-2010,BoMi-Science-2012}.
Second, the electron-electron interactions become more dominant leading to non-stationary hole states which require a many-body description of the ionic system~\cite{WeSc-JPC-1996,BrCe-JCP-2003,NeRe_NJoP10-2008,KuLu-JPCA-2010}.

Multiorbital contributions in atoms due to a complex recombination step involving more than one electron have been experimentally seen~\cite{ShVi-NatPhys-2011}.
Even though direct contributions from orbitals other than the outermost $p$ shell do not exist (in contrast to molecules),
many-body interchannel coupling can, however, access deeper shells and make them 'indirectly' contribute.
This has been theoretically shown for the $3p$ and $3s$ channels in argon~\cite{BrHu-PRL-2012} and the $4d$ and $5p$ channels in xenon~\cite{PaSa-PRL-2013}.
In recent HHG experiments with mixed gases~\cite{KaTa-PRL-2007}, constructive and destructive interference of the generated HHG light from different atomic species has been observed. This interference is not due to multiorbital effects but due to different intrinsic dipole phases of the atomic species.

Also a hole motion can be triggered via tunnel ionization within the outermost $p$ shell.
In recent experiments~\cite{GoKr-Nature-2010,WiGo-Science-2011} the formation of a hole wavepacket involving the spin-orbit-split $4p_{3/2}$ and $4p_{1/2}$ orbitals in krypton was demonstrated.
It is even possible to measure the degree of coherence of the hole wavepacket via attosecond transient absorption spectroscopy~\cite{SaYa-PRA-2011}.
The influence of such a spin-orbit driven hole motion on HHG has, however, not been studied so far.
Since the HHG process itself is spin-insensitive, one might even think that spin-orbit effects do not influence the HHG process.
 
Here, we show theoretically that spin-orbit-driven hole motion, which is a multiorbital effect, does influence HHG.
To demonstrate this, we solve the many-body Schr\"odinger equation by the time-dependent configuration-interaction singles (CIS) approach~\cite{GrSa-PRA-2010,PaSy-PRA-2012}, which describes the entire $N$-electron wavefunction and, consequently, all atomic shells that may or may not contribute in the HHG spectrum.
Specifically, we show that the HHG yield depends on the normalized time $t_c/T_\textrm{so}$, which measures the time the electron spends in the continuum, $t_c$, in units of the spin-orbit period $T_\textrm{so}$.
When the hole motion is purely due to the spin-orbit effect, the angular distribution of the HHG radiation is insensitive to the hole dynamics.
Also the detailed structure of the spectrum depends only slightly on $t_c/T_\textrm{so}$ because the radial recombination matrix elements, which responsible for the details of the HHG spectrum, do not change.
What depends mostly on spin-orbit effects is, however, the overall HHG yield.

The time the electron spends in the continuum, $t_c = t_r - t_i$, is linearly dependent on the wavelength of the NIR field $\lambda$.
Here $t_i$ is the time of ionization, which starts the hole motion, and $t_r$ is the time of recombination, which stops the hole motion.
By changing the wavelength the time at which the electron recombines and the hole state is probed can be controlled.
In general, varying the spin-orbit period $T_\textrm{so}$ is also sufficient to achieve the same effect.
Unfortunately, $T_\textrm{so}$ is an intrinsic property of the atom and experimentally only $\lambda$ is tunable. 

The further discussion is structured as follows:
In Sec.~\ref{s2.1} basic aspects of our many-body TDCIS approach are explained, which we use to study the multiorbital HHG process.
In Sec.~\ref{s2.2} we discuss in more detail the spin-orbit coupling and how it can be treated as a perturbation.
In Sec.~\ref{s2.3} the consequences of the purely spin-orbit-driven hole motion for HHG are discussed.
The numerical results are presented in Sec.~\ref{s3} for atomic krypton.
We show the spin-orbit dependence of the HHG yield by varying $T_\textrm{so}$ (see Sec.~\ref{s3.1}) and $\lambda$ (see Sec.~\ref{s3.2}).
Furthermore, we identify pulse parameters where the spin-orbit effect leads to an increase in the HHG yield by increasing $\lambda$. 

Atomic units are employed throughout unless otherwise
indicated.

\section{Theory}
\label{s2}

\subsection{TDCIS} 
\label{s2.1}

Our implementation of the TDCIS approach~\cite{GrSa-PRA-2010} and the extension to spin-orbit interactions for the occupied orbitals~\cite{PaSy-PRA-2012}, which we use here~\cite{n.xcid} to solve the $N$-body Schr\"odinger equation, has been described in previous publications.
We have already successfully applied our TDCIS approach to a wide spectrum of processes ranging from attosecond multiorbital phenomena~\cite{PaSa-PRL-2011,PaSy-PRA-2012} to nonlinear x-ray ionization~\cite{SyPa-PRA-2012} and to strong-field physics including non-adiabatic tunnel ionization~\cite{KaPa-PRA-2013} and multi-orbital HHG processes~\cite{PaGr-PRA-2012,PaSa-PRL-2013}.

The TDCIS wave function ansatz reads~\cite{GrSa-PRA-2010}
\begin{eqnarray}
  \label{eq:tdcis}
  \ket{\Psi(t)}
  &=&
  \alpha_0(t) \, \ket{\Phi_0}
  +
  \sum_{a,i}
    \alpha^a_i(t) \, \ket{\Phi^a_i}
  ,
\end{eqnarray}
where $\ket{\Phi_0}$ is the Hartree-Fock ground state and $\ket{\Phi^a_i}= \hat c^\dagger_a \hat c_i \ket{\Phi_0}$ are singly excited configurations with an electron removed from the initially occupied orbital $i$ and placed in the initially unoccupied orbital $a$.
Due to this ansatz the electron can be removed from any orbital and not just from the outermost orbital as usually done in the single-active electron approximation~\cite{HiFa-PRA-2011,ChWu-PRA-2013}.
Furthermore, the hole state (represented by the index $i$) and the excited/ionized electron (represented by the index $a$) can move independently making TDCIS an effective two-active particle theory that goes beyond the independent particle picture~\cite{SzOs-book}.
The interaction between these two effective particles due to electron-electron interaction~\cite{PaSa-PRL-2011,PaSa-PRL-2013} or light-matter interaction~\cite{PaSy-PRA-2012} leads to many surprising phenomena.

The resulting equations of motion for the CIS coefficients read
\begin{subequations}
\begin{align}
  \label{eq:eoms.1}
  i\partial_t \, \alpha_0(t)
  &=
  -E(t)\, \sum_{a,i} \sbra{\Phi_0} \hat z \sket{\Phi^a_i}
  \\\nonumber
  \label{eq:eoms.2}
  i\partial_t \, \alpha^a_i(t)
  &= 
  \sbra{\Phi^a_i} \hat H_0 \sket{\Phi^a_i} \, \alpha^a_i(t) 
  +
  \sum_{b,j}
    \sbra{\Phi^a_i} \hat H_1 \sket{\Phi^b_j}
    \alpha^b_j(t)
  \\ &
  -E(t)
  \Big(\!
    \sbra{\Phi^a_i} \!\hat z \! \sket{\Phi_0}
    \alpha_0(t)
    +\!
    \sum_{b,j}
      \sbra{\Phi^a_i} \!\hat z\! \sket{\Phi^b_j}
      \alpha^b_j(t)
  \Big)
  ,
\end{align}
\end{subequations}
where $\hat H_0= \sum_n \left[ \frac{\hat{\bf p}^2_n}{2}  - \frac{Z}{|\hat{\bf r}_n|} + V_\textrm{MF}(\hat{\bf r}_n) \right] - E_\textrm{HF}$ includes all one-particle operators (kinetic energy, attractive nuclear potential, and the the mean-field potential $\hat V_\textrm{MF}$).
The entire energy spectrum is also shifted down by the Hartree-Fock ground state energy, $E_\textrm{HF}$, for convenience.
The nuclear charge is given by $Z$ and the index $n$ runs over all $N$ electrons in the system.
The light-matter interaction for linearly polarized pulses is given in the dipole approximation by $-E(t)\, \hat z$ with $\hat z = \sum_n \hat z_n$, where $E(t)$ is the electric field of the pulse.
All the electron-electron interactions that cannot be described by the mean-field potential $\hat V_\textrm{MF}$ are captured by $\hat H_1 = \frac{1}{2}\sum_{n,n'} \frac{1}{|\hat{\bf r}_n - \hat{\bf r}_{n'}|} - \sum_n \hat V_\textrm{MF}(\hat{\bf r}_n)$~\cite{RoSa-PRA-2006}. 
Since we employ a complex absorbing potential~\cite{RiMe-JPB-1993} to eliminate the outgoing electron wavepacket when it reaches the end of the numerical grid, we have to use the symmetric inner product $\left(\cdot\right|,\left|\cdot\right)$ instead of the hermitian one $\left<\cdot\right|,\left|\cdot\right>$.

\subsection{Spin-Orbit} 
\label{s2.2}

\begin{figure}[t!]
  \centering  
  \rmpdfinfo
  \includegraphics[clip,width=\linewidth]{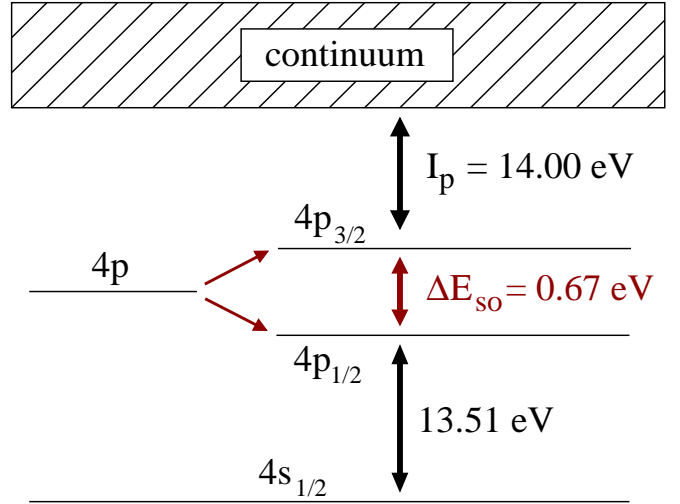}
  \caption{(color online) 
    Spin-orbit coupling within the outer-most $4p$ shell of atomic krypton.
    The energy splitting is relatively small compared to the ionization potential $I_p$.
  }
  \label{f1}
\end{figure}

As spin-orbit interaction is usually a small effect in the outer-most $p$-shell of noble gas atoms (see Fig.~\ref{f1}) leading to energy gaps (up to 3.8~eV in radon~\cite{NIST_website}), the spin-orbit-coupled orbitals are well described in terms of first-order degenerate perturbation theory.
At the end of this section, we quantify how good this approximation is in terms of dipole transition matrix elements.
But first, we review how we include in TDCIS spin-orbit coupling for the occupied orbitals~\cite{RoSa-PRA-2009}.
The spin-orbit interaction in the virtual (Rydberg + continuum) orbitals are ignored, since it is even smaller than for the occupied orbitals as we will see in the following.

Let us consider the non-relativistic limit of the spin-orbit interaction
~\cite{LaLi-QM-book,Jackson-book,Fa-AJP-1967}
\begin{eqnarray}
  \label{eq:H_so}
  \hat H_\textrm{so}
  &=&
  \frac{\alpha^2}{2}\, \frac{1}{r}\frac{dV}{dr} \sum_n \hat{\bf l}_n \!\cdot \hat{\bf s}_n
\end{eqnarray}
to understand the rapid decrease in the importance of the spin-orbit interaction with increasing orbital energy.
Here, $\hat{\bf l}_n$ and $\hat{\bf s}_n$ are the orbital angular momentum operator and the spin operator of the $n^\textrm{th}$ electron.
The potential $V$ denotes the mean-field potential plus the nuclear Coulomb potential.
For hydrogen, $V$ includes only the nuclear Coulomb potential and one finds $\hat H_\textrm{so} \propto r^{-3}$
~\cite{Fa-AJP-1967,CoTa_QM2_book}.
Note that Eq.~\eqref{eq:H_so} is an approximative extension (from a one-electron atom) to a many-electron atom neglecting explicit two-body spin-orbit terms
~\cite{BlWa-PRSA-1962,BlWa-PRSA-1963}.

For many-electron systems, the mean-field potential $V$ decreases as $-Z/r$ for small radii and as $-1/r$ for large radii (for occupied orbitals; for unoccupied orbitals the decrease is exponential for large $r$).
In both cases the radial dependence of the spin-orbit interaction goes as $r^{-1}\,\partial_r V(r) \propto r^{-3}$.
Applying this result to Rydberg and continuum states, which are quite delocalized and on average far away from the nucleus, we find that the strength of the spin-orbit interaction is strongly reduced in comparison to occupied orbitals and, therefore, can be neglected.
As a result, the virtual orbitals, $\ket{a}=\ket{n_a,l_a,m^L_a;m^S_a}$, can be directly taken from the non-relativistic Hartree-Fock calculations, where $n_a$ stands for the principal  quantum number of orbital $a$, $l_a$ is the orbital angular momentum, $m^L_a$ is its projection on the laser polarization axis, and $m^S_a$ is the spin projection.
Since the spin of all electrons is always $s_i=\half$, the spin quantum number is omitted in characterizing the orbital.

After rewriting the spin-orbit operator $2\,\hat{\bf l}\cdot\hat{\bf s} = \hat{\bf j}^2-\hat{\bf l}^2-\hat{\bf s}^2$ in terms of the total angular momentum $\hat{\bf j}$, it becomes clear that the new orbitals are eigenstates of the operators $\hat{\bf j}^2,\hat{\bf l}^2,\hat{\bf s}^2$, and $\hat{j}_z$, which constitute the coupled LS-basis.
The new spin-orbit-coupled orbitals $i$ expressed in terms of the old orbitals (without spin-orbit coupling) read
\begin{align}
  \label{eq:wfct_so}
  \ket{i}
  :=&
  \ket{n_i,l_i,j_i,m^J_i}
  \\\nonumber
  =&
  \sum_{m^S_i,m^L_i}
    C^{j_i,m^J_i}_{l_i,m^L_i;s_i,m^S_i} \ket{n_i,l_i,m^L_i;m^S_i}
  ,
\end{align}
where the Clebsch-Gordan coefficient is given by $C^{l_3,m_3}_{l_1,m_1;l_2,m_2}=\braket{l_1,m_1;l_2,m_2}{l_3,m_3}$, and $m^J_i,m^L_i,$ and $m^S_i$ are the projections of the total angular momentum, orbital angular momentum, and of the spin on the laser polarization axis, respectively.

Within perturbation theory the change in the angular distribution of the new occupied orbitals can be fully explained by angular momentum coupling of $l_i$ and $s_i$.
The radial part of the orbital wavefunction, which depends only on $n$ and $l$, does not change.
This is, however, only true in the perturbative limit.
In a truly relativistic Dirac-Fock (DF) calculation the radial wavefunction depends also on $j$.

\begin{table}[ht!]
  \caption{\label{tab:1} 
    The reduced dipole moments $\abs{\redmom{4p_j}{r}{3d_{j'}}}$ of the ionic transitions $3d_{j'}^{-1} \rightarrow 4p_j^{-1}$.
    The results obtained via perturbation theory are compared with literature values~\cite{SaYa-PRA-2011}, which have been calculated by using relativistic Dirac-Fock calculations performed with {\sc grasp}~\cite{GRASP-DyGr-CPC-1989}.
    The perturbative results are calculated with Eq.~\eqref{eq:dipole} and with the non-relativistic reduced matrix element $\abs{\redmom{3d}{r}{4p}}=0.298$ obtained from the {\sc xcid} code. Values are given in atomic units.
  }

  \begin{tabular}{ c | c | c | c}
    \hline
    transition  & perturbation theory & Ref.~\cite{SaYa-PRA-2011} & error \\
    \hline
    $\abs{\redmom{4p_{3/2}}{r}{3d_{5/2}}}$ & 0.326 & 0.345 & -5\%  \\
    $\abs{\redmom{4p_{3/2}}{r}{3d_{3/2}}}$ & 0.108 & 0.112 & -3\%  \\
    $\abs{\redmom{4p_{1/2}}{r}{3d_{3/2}}}$ & 0.243 & 0.264 & -7\%  \\
    \hline
  \end{tabular}
\end{table}

To give an estimate of how good our orbitals (obtained from perturbation theory) are in comparison with fully relativistic DF orbitals, we compare in Table~\ref{tab:1} reduced dipole transition strengths between the $3d$ and $4p$ shells of krypton (which is also the atom of interest in Sec.~\ref{s3}).
Note that in CIS the transition elements between the ionic ($N-1$ particle) states $i^{-1}$ and $j^{-1}$ reduces to one-particle transition elements $\redmom{i}{r}{j}$ due to Koopmans' theorem~\cite{SzOs-book}, where $i^{-1}$ stands for a configuration with one electron missing in the orbital $i$ with respect to the neutral Hartree-Fock ground state.
As Table~\ref{tab:1} shows, the transition strengths obtained via degenerate perturbation theory~\cite{Va-book}
%
\begin{align}
  \label{eq:dipole}
  \redmom{n',l',j'}{r}{n,l,j}
  &=
  \redmom{n',l'}{r}{n,l}    
  \\\nonumber 
  & \hskip-12ex \times
  (-1)^{j+l'+s+1}
  \sqrt{(2j+1)(2j'+1)}
  \wigsixj{l}{s}{j}{j'}{1}{l'}
  ,
\end{align}
deviate less than 10\% from the relativistic {\sc grasp} calculations~\cite{GRASP-DyGr-CPC-1989} confirming the applicability of perturbation theory.
The expression $\{ \cdot\}$ is the Wigner-6j symbol~\cite{Zare-book} and $s=\half$ is the spin of the electron.

Finally, we only need to find the orbital energies of the new spin-orbit-coupled orbitals.
Instead of determining them with perturbation theory, we set the orbital energies to the experimental values.
This has the advantage that we have the exact ionization potential, which is important for accurate tunnel ionization dynamics.

\subsection{HHG} 
\label{s2.3}

In Sec.~\ref{s2.3.1}, we derive an expression for the spin-orbit dependence of the HHG yield based on the semi-classical three-step model~\cite{Co-PRL-1993,ScKu-PRL-1993}.
In Sec.~\ref{s2.3.2}, we give a short argumentation why the HHG angular distribution is not affected by the spin-orbit coupling.

\subsubsection{HHG Yield}
\label{s2.3.1}
We will ignore interchannel effects, which would entangle the photoelectron with the ionic state.
Without this entanglement, the overall state can be written as a product of photoelectron and hole wavefunctions.

From a time-dependent quantum calculation with a classical description of the laser pulse, the HHG spectrum $S(\omega)$ can be calculated via the expectation value of the dipole acceleration~\cite{Jackson-book}, which can be expressed in three major ways using the dipole operator, the momentum operator, or the dipole acceleration operator~\cite{Pa-EPJST-2013}. 
Here, we choose the the length form involving the dipole operator:
\begin{eqnarray}
  \label{eq:hhg}
  S(\omega)  
  &\propto& 
  \abs{
    \int\! dt\ e^{-i\omega t} \left[\partial^2_t \left< z\right>(t)\right] 
  }^2
  .
\end{eqnarray}
The dipole expectation value
\begin{eqnarray}
  \label{eq:ev}
  \left< z\right>(t)
  &=&
  \sum_{abi} \,
    [\alpha^b_i(t)]^* \alpha^a_i(t) \bra{\Phi^b_i}\hat z\ket{\Phi^a_i}
  \\\nonumber&&+
  \sum_{aij} \,
    [\alpha^a_j(t)]^* \alpha^a_i(t) \bra{\Phi^a_j}\hat z\ket{\Phi^a_i}
  \\\nonumber&&+
  \sum_{ai} \, \left(
    \alpha^*_0(t) \alpha^a_i(t) \bra{\Phi_0}\hat z\ket{\Phi^a_i}
  + c.c.
    \right)
  ,    
\end{eqnarray}
captures the continuum motion of the ionized electron from all ionization pathways [first term in Eq.~\eqref{eq:ev}] and the motion of the ionic hole [second term in Eq.~\eqref{eq:ev}].
The HHG spectrum is, however, primarily due to the third term in Eq.~\eqref{eq:ev}, which describes the recombination of the photoelectron with the ionic hole.

In the further discussion, we will dissect the HHG mechanism and focus on how spin-orbit coupling affects the last term of Eq.~\eqref{eq:ev} and, consequently, the HHG emission.
To do so, it is convenient to use the semiclassical three-step picture~\cite{Co-PRL-1993,ScKu-PRL-1993}, where step 1 is the tunnel ionization of the electron, step 2 is the field-driven motion of the electron in the continuum, and step 3 is the recombination of the electron with the ion.
The transition matrix elements appearing in step 3 are the same matrix elements that appear in photoionization~\cite{MoLi-PRL-2008}.
As we have shown in previous works~\cite{PaSa-PRL-2011,KrPa-submitted}, TDCIS is able to capture multiorbital and many-body physics in photoionization.


\paragraph{Tunnel ionization} affects predominantly the outermost $p$ orbital aligned in the direction of the linearly polarized light; from here on referred to as $np_0$.
Note that when $np_0$ is used it stands for the uncoupled $p$ orbital with $m^L=0$.
The expression $np_j$ stands for the spin-orbit coupled $p$ orbital with total angular momentum $j$; no statement is made here about any angular momentum projection.
The initial, ionized part of the wavefunction at the time of tunnel ionization, $t_i$, can be written as a product,
\begin{eqnarray}
  \label{eq:wfct_step1}
  \ket{\Psi(t_i)}
  &=&
  \sum_{m^S_a}
  \underbrace{
    \ket{\chi_{m^S_a}(t_i)
    \otimes
    \kappa_{m^S_a}(t_i)
    }    
  }_{\ds \sum_{a}  \alpha^a_{np_0,m^S_a}(t) \ket{\Phi^{a}_{np_0;m^S_a}}}
  ,
\end{eqnarray}
of an ionic part $\ket{\kappa_{m^S_a}(t_i)}=\ket{\Phi_{np_0;m^S_a}}$ and a photoelectron part $\ket{\chi_{m^S_a}(t_i)}=\sum_{a} \alpha^{a}_{4p_0;m^S_a}(t_i)\ket{a}$.
Here, $\ket{\Phi_i}=\hat c^\dagger_i \ket{\Phi_0}$ is an $(N-1)$-electron state characterizing the ion.
Equation~\eqref{eq:wfct_step1} relies strongly on the fact that the spin-orbit splitting, $\Delta E_\textrm{so}$, is small compared to the ionization potential $I_p$.
If the difference in the ionization potential is large, the outermost $np_{3/2}$ orbital is dominantly selected and the $np_0$ character will reduce.
In the discussion that follows we will see the benefit of the introduced separation between photoelectron and the ionic wavefunctions.
Note that the spatial distribution of the photoelectron wavefunction for a spin-up and spin-down electron is the same, $\ket{\chi_{m^S_a}(t_i)}=\ket{\chi_{-m^S_a}(t_i)}$.

\paragraph{Field-driven propagation} affects only the photoelectron $\ket{\chi_{m^S_a}(t_i)}$ (action is independent of ${m^S_a}$).
Due to the spatial separation, $\ket{\chi_{m^S_a}(t)}$ evolves independently of $\ket{\kappa_{m^S_a}(t)}$~\cite{PaSy-PRA-2012} and is identical to the non-relativistic case without spin-orbit coupling.
The hole state propagates (basically) under field-free conditions~\footnote{Ionic polarization does exist but the overall effect (compared to the spin-orbit interaction) is small~\cite{PaSy-PRA-2012} and can be ignored in our current discussion.
The numerical results in Sec.~\ref{s3} account for some polarization effects (within the 1-hole configurations) but they do not matter.}.
Since we include spin-orbit coupling, the hole propagation is preferably described in the coupled basis.
The overall $N$-electron state during step 2 can be still written as a product of electron and hole states,
\begin{align}
  \label{eq:wfct_step2}
  \ket{\Psi(t)}
  &=
  \sum_{m^S_a}
    \ket{\chi_{m^S_a}(t)}
    \otimes
    \ket{\kappa_{m^S_a}(t)}
  ,
  \\
    \label{eq:hole_step2}
  \ket{\kappa_{m^S_a}(t)}
  &=
  \sum_j
    C^{j,\sigma}_{1,0;s,{m^S_a}} \,
    e^{i\varepsilon_j\,(t-t_i)}
    \ket{np^{m^J=\sigma}_j} 
  \\\nonumber
  &=
  \sum_{j,m^L}
    C^{j,\sigma}_{1,0;s,{m^S_a}} \,
    C^{j,\sigma}_{1,m^L;s,{m^S_a}-m^L} \,
    e^{i\varepsilon_j\,(t-t_i)}
  \\\nonumber
  &\quad\qquad \times
    \ket{np_{m^L};m^S_a-m^L}
  ,
\end{align}
where $\varepsilon_j$ is the energy of the $np_j$ orbital with the initial condition given in Eq.~\eqref{eq:wfct_step1}.

\paragraph{Recombination} is again a spin-insensitive process and one finds that the photoelectron can only recombine with the $np_0^{-1}$ hole state (neglecting interchannel effects~\cite{BrHu-PRL-2012,PaSa-PRL-2013}).
But due to the spin-orbit coupling during the field-driven propagation (step 2), there is a final spin-orbit effect on the recombination strength due to the spin-orbit-driven hole motion.
Inserting Eq.~\eqref{eq:wfct_step2} into Eq.~\eqref{eq:ev} and focusing only on the last (HHG relevant) term in Eq.~\eqref{eq:ev}, we find the dipole moment 
\begin{eqnarray}
  \label{eq:dipole_so}
  \left< z\right>(t)
  &=&   
   \sum_{m^S_a}
     \bra{\Phi_0} \hat z \ket{\chi_{m^S_a}(t) \otimes \kappa_{m^S_a}(t)}
     + c.c.
  \\\nonumber
  &=&
  2\frac{
    2 + e^{-i\Delta E_\textrm{so}\,(t-t_i)}
  }{3}
  \,e^{-iI_p\,(t-t_i)} \!
  \bra{np_0} \!\hat z\! \ket{\chi(t)}
  \!+\! c.c.
\end{eqnarray}
can be separated in three terms:
(1) a term depending on the spin-orbit splitting $\Delta E_\textrm{so} = \varepsilon_{3/2} - \varepsilon_{1/2}$,
(2) a phase term due to the ionization potential $I_p=-\varepsilon_{3/2}$ of the outer-most orbital,
and (3) the spin-independent recombination transition element $\bra{np_0} \hat z \ket{\chi(t)}$.
The last two terms (including the overall factor 2) are identical to the expression obtained without spin-orbit interaction (up to a slightly different $I_p$).
In the last step of Eq.~\eqref{eq:dipole_so} we drop the spin index of the photoelectron, since the spatial distribution and the matrix element does not depend on it.
The entire spin-orbit dependence in the dipole moment comes from the additional term, $(2 + e^{-i\Delta E_\textrm{so}\,(t-t_i)})/3$, which reflects the spin-orbit beating of the hole during the field-driven propagation discussed previously.

The emission of photons is due to fast temporal changes in $\left< z\right>(t)$ around the time of recombination, $t \approx t_r$.
The classical trajectory associated with the maximum returning kinetic energy generates photon energies around the cut-off energy, $\omega_\textrm{cut-off}$.
The time of recombination for this trajectory is $t_c = 0.67~T_\lambda$, where $T_\lambda$ is the period of one cycle of the driving IR pulse.
For $\lambda=800$~nm, the period is $T_\lambda=2.67$~fs and the time of recombination is $t_c=1.79$~fs.

The spin-orbit beating happens on a much slower time scale than the duration of the recombination event, $T_\textrm{so}=2\pi/\Delta E_\textrm{so}  \gg 2\pi/\omega_\textrm{cut-off}\lesssim 100$~as, such that the ionic hole state and, hence, the prefactor $c_\textrm{so}(t_c)=(2 + e^{-i\Delta E_\textrm{so}\,t_c})/3$ can be viewed as constant during the recombination event.
Consequently, the HHG yield near the cut-off energy depends parametrically on $t_c$.
The HHG yield near the cut-off energy, $S_\textrm{so}(\omega_\textrm{cut-off})$, is scaled by $\left|c_\textrm{so}(t_c)\right|^2$ with respect to the non-spin-orbit single-channel result, $S_\textrm{no-so} (\omega_\textrm{cut-off})$,
\begin{align}
  \label{eq:hhg_so}
  S_\textrm{so}(\omega_\textrm{cut-off})
  &=
  \frac{5+4\,\cos(\Delta E_\textrm{so}\,t_c)}{9}
  S_\textrm{no-so} (\omega_\textrm{cut-off})
  .
\end{align}

In this entire discussion, we assumed that the difference in the ionization potentials can be ignored such that a perfect $np_0$ hole wavepacket is created at $t=t_i$.
However, even though the difference in $I_p$ is small it will lead to a preferred ionization out of the outermost $np_{3/2}$ orbital exceeding the 2:1 preference for $np_{3/2}$, which already exists due to the angular momentum coupling [see Eq.~\eqref{eq:dipole_so}]. 
The increased preference for one of the two spin-orbit-split orbitals leads to a reduced contrast in the spin-orbit prefactor $\left|c_\textrm{so}(t_c)\right|^2$.
Therefore, Eq.~\eqref{eq:hhg_so} presents an upper bound for the beating contrast due to spin-orbit coupling.

The general structure of the HHG will not be affected by the spin-orbit beating as it becomes apparent from Eq.~\eqref{eq:hhg_so}.
The spin-orbit motion affects only the yield. 
The details of the HHG spectrum originate from the recombination elements $\bra{np_0} \hat z \ket{\chi(t)}$ [see Eq.~\eqref{eq:dipole_so}], and these are the same wether or not spin-orbit interactions are present.

Furthermore, we focus here on the cut-off region of the HHG spectrum, which in a classical picture originates only from one specify trajectory.
For other energy regions of the HHG spectrum this is not true anymore and a unique $t_c$ cannot be identified.
This has the consequence that each part of the HHG spectrum scales a bit differently leading to a spectrum that is not just scaled by a common global prefactor.
By choosing the cut-off region, we avoid this additional complication and we can assign a unique $t_c$.

\subsubsection{Angular HHG distribution}
\label{s2.3.2}
We have seen that the overall HHG yield depends on the spin-orbit splitting.
The question remains whether also the angular distribution depends on spin-orbit effects.
A change in the angular distribution would be reflected in non-zero dipole moments $\left< x\right>(t)$ and $\left< y\right>(t)$.
Without spin-orbit effects these moments are zero, since the photoelectron and the hole have always the same $m^L_a=m^L_i$ character and only $\Delta m^L=0$ transitions can occur.
In other words, the total $N$-electron state $\ket{\Psi}$ has always the overall orbital angular momentum $M^L=\sum_n m^L_n=m^L_a-m^L_i=0$, since the initial HF ground state has $M^L=0$ and the linearly polarized pulse can not change $M^L$.

With spin-orbit coupling, one finds the $m^L_i$ character of the hole changes by $\pm 1$ while the $m^L_a=0$ character of the photoelectron remains untouched.
Naively one would think $\Delta m^L=\pm1$ is now possible in the recombination step and, hence, the dipole moments $\left< x\right>(t)$ and $\left< y\right>(t)$ would become non-zero.
This is, however, not true. 
The spin-orbit coupling changes $m^L_i$ but at the same time it also changes the spin $m^S_i$ of the hole in the opposite direction such that the sum is always the same, i.e., $m^J_i=m^L_i+m^S_i$.
Since recombination is a dipole transition, where no spin flip can occur, the photoelectron can only recombine with hole states that have the same spin, $m^S_i=m^S_a$ and consequently the same orbital angular momentum, $m^L_a=m^L_i$.

In terms of the overall $N$-electron states the argument is similar.
With spin-orbit coupling the characters $M^L$ and $M^S$ are not conserved anymore but $M^J=\sum_n m^J_n=0$ is conserved.
Due to the Wigner-Eckart theorem~\cite{Zare-book}, only dipole transitions in $z$ direction can occur.
Dipole transitions in $x$ or $y$ directions change $M^J$ by $\pm 1$.

\section{Results}
\label{s3}

In this section, we present numerical results~\footnote{
For the system radius if $200~a_0$ we used 1000 grid points, which are non-uniformly distributed~\cite{GrSa-PRA-2010} characterized by the mapping parameter of $\zeta=3$.
The complex absorbing potential starts at $180~a_0$ and has a strength of $5\cdot 10^{-3}$.
Orbitals with an energy larger than 15~a.u.are omitted resulting into around 350 states per angular momentum.
The maximum angular momentum is 120 and for orbital angular momenta $l>5$ the residual Coulomb between electron and ion is reduced to the monopole term resulting in a one-particle $-1/r$ potential.
}
based on the TDCIS approach for the spin-orbit dependence of the HHG yield (as explained in Sec.~\ref{s2.3.1}).
The atom of choice is krypton with a spin-orbit splitting of 0.67~eV between the orbitals $4p_{3/2}$ and $4p_{1/2}$, and an ionization potential of the outermost $4p_{3/2}$ of $|\varepsilon_{3/2}|=14.0$~eV~\cite{NIST_website} (see Fig.~\ref{f1}).
All results presented are based on TDCIS calculations with the {\sc xcid} code~\cite{n.xcid}, where only the orbitals in the $4p$ manifold are active and interchannel coupling among these orbitals is included (even though interchannel coupling is not important in this discussion).
The $4s$ and the $3d$ shells are frozen and do not affect the presented results. 
Calculations including these two shells are presented in \ref{app1}.

The influence of the spin-orbit dynamics of the ion hole on the HHG spectrum can be solely expressed by the quantity $t_c/T_\textrm{so}$ as derived in Sec.~\ref{s2}.
To demonstrate this effect, either the time the electron spends in the continuum, $t_c$, or the spin-orbit period, $T_\textrm{so}$, may be varied.
In Sec.~\ref{s3.1} we vary $T_\textrm{so}=2\pi/\Delta E_\textrm{so}$ by changing the spin-orbit splitting $\Delta E_\textrm{so}$.
In Sec.~\ref{s3.2} we vary $t_c$ by changing the wavelength of the driving NIR pulse.
This case is experimentally more realistic but the identification of the spin-orbit effect is more challenging because the spin-orbit-free HHG spectrum changes also with $\lambda$~\cite{TaAu-PRL-2007,FoMa-PRL-2008,ShTr-PRL-2009}.

\subsection{Varying Spin-Orbit Splitting}
\label{s3.1}

First, we demonstrate the spin-orbit effect by keeping the driving pulse fixed and only changing the spin-orbit splitting $\Delta E_\textrm{so}=\varepsilon_{3/2}-\varepsilon_{1/2}$.
This can be easily done in theory, since $\varepsilon_i$ are input parameters.
To do so, we keep the ionization potential $I_p=-\varepsilon_{3/2}$ of the outermost valence shell fixed and only change the ionization potential of the energetically lower-lying $4p_{1/2}$ shell,
\begin{eqnarray}
  \label{eq:4p12}
  \varepsilon_{1/2}
  &:=&
  \varepsilon_{3/2}
  -
  \Delta E_\textrm{so}
  .
\end{eqnarray}
This has the advantage that the shape and the strength of the HHG spectra for different $\Delta E_\textrm{so}$ are comparable with each other, since the tunneling rate out of $4p_{3/2}$ is always the same.
A variation in $\varepsilon_{3/2}$ would strongly change the tunnel ionization rate and the overall HHG yield, which would make it harder to uniquely identify the influence of the spin-orbit beating of the hole state.

\begin{figure}[t!]
  \centering  
  \rmpdfinfo
  \includegraphics[clip,width=\linewidth]{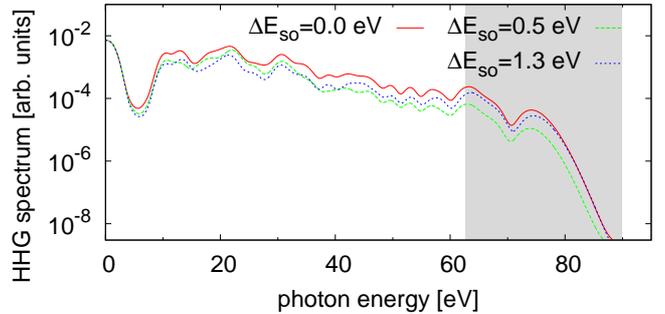}
  \caption{(color online) 
    The HHG spectrum for spin-orbit splittings of $\Delta E_\textrm{so}=0.0$~eV (solid-red), $\Delta E_\textrm{so}=0.5$~eV (dashed green), and $\Delta E_\textrm{so}=1.3$~eV (dotted blue).
    The highlighted area around the cut-off energy is used to discuss the spin-orbit influence on the HHG yield.
    The driving pulse has a cycle-averaged peak intensity of $8.8\times 10^{13}$~W/cm$^2$, FWHM-duration (of the intensity) of 10~fs, and a wavelength of 1500~nm.
  }
  \label{f2.0}
\end{figure}

In Fig.~\ref{f2.0} HHG spectra~\footnote{
Note that the HHG spectra are convolved with a 2.7~eV wide Gaussian for a smoother appearance.
}
are shown for different spin-orbit strengths but for the same driving pulse (for pulse parameters see figure caption).
The HHG yield looks almost unchanged for low energies. 
At high energies close to the cut-off energy, one finds that monotonically changing the spin-orbit splitting does not change the HHG yield monotonically. 
For large spin-orbit splitting of $\Delta E_\textrm{so}=1.3$~eV (blue-dotted line) the yield is larger than for $\Delta E_\textrm{so}=0.5$~eV (green-dashed line).
But both HHG yields are smaller than for the non-spin-orbit case (red-solid line).
Consequently, having spin-orbit will only lead to a reduction in the overall yield compared to having no spin-orbit interaction [see Eq.~\eqref{eq:hhg_so}].
But increasing the spin-orbit interaction does not mean a monotonic change.

To derive Eq.~\eqref{eq:hhg_so} we made the assumption that the HHG emission is generated in a time interval that is short compared to $T_\textrm{so}$.
This is, however, not true for the entire HHG spectrum.
But for specific energy regions, especially near the cut-off, this assumption holds, because only one electron trajectory contributes to this specific energy region.
Therefore, the HHG yields shown in the following are integrated quantities from $\omega_\textrm{cut-off}-E_h/2$ to $\omega_\textrm{cut-off}+E_h/2$ (highlighted area in Fig.~\ref{f2.0}), where $\omega_\textrm{cut-off}=I_p+3.17\,U_p$ is the HHG cut-off energy and $E_h=27.21$~eV ($E_h=1$~a.u.) is the Hartree energy\footnote{Also a relative energy window (e.g., from $90\% E_\textrm{cut-off}$ to $110\% E_\textrm{cut-off}$) could have been used for the integration. The qualitative features of the curves discussed do not change when using a energy window with a fixed relative width instead of a fixed absolute width.}.
Another advantage of choosing the cut-off region is that multiple-path interference does not exist in the cut-off region~\cite{ShIs-PRL-2007} .

\begin{figure}[t!]
  \centering  
  \rmpdfinfo
  \includegraphics[clip,width=\linewidth]{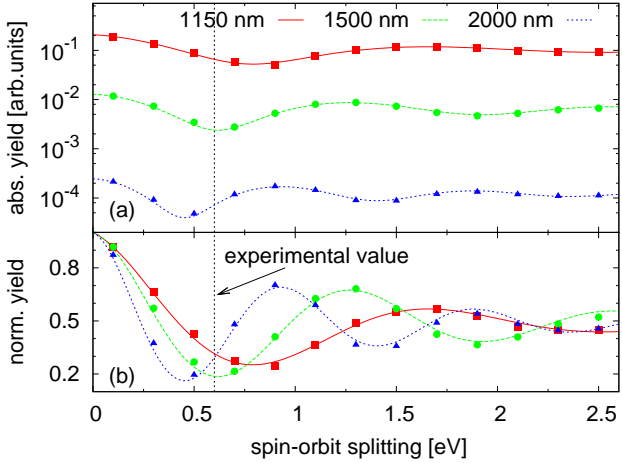}
  \caption{(color online) 
    The (a) absolute and (b) the normalized HHG yield around the HHG cut-off energy as a function of the spin-orbit splitting $\Delta E_\textrm{so}$ for the wavelengths $\lambda=1150$~nm, 1500~nm, and 2000~nm.
    The points represent the TDCIS results.
    The lines are fits to the TDCIS data using the function defined in Eq.~\ref{eq:fit_fct}.
    The driving pulse has a cycle-averaged peak intensity of $8.8\times 10^{13}$~W/cm$^2$ and FWHM-duration (of the intensity) of 10~fs.
    The experimentally determined spin-orbit splitting at 0.67~eV is highlighted in both panels.
  }
  \label{f2}
\end{figure}

In Fig.~\ref{f2} the absolute (a) and the normalized (b) HHG yields are shown for the NIR wavelengths $\lambda=1150$~nm, 1500~nm, and 2000~nm.
The yields in Fig.~\ref{f2}(b) are normalized with respect to the $\Delta E_\textrm{so}=0$~eV point, which corresponds no spin-orbit coupling.

The absolute yield decreases rapidly with the wavelength as expected~\cite{TaAu-PRL-2007}. 
The oscillations in the yield can be seen at all three wavelengths and are due to the spin-orbit beating.
Particularly in the normalized yield the oscillations are clearly visible.
The points in Fig.~\ref{f2} are results from the TDCIS calculations whereas the lines are fits to the TDCIS data using the fitting function 
\begin{eqnarray}
  \label{eq:fit_fct}
  f_\textrm{fit}(\Delta E_\textrm{so})
  &=&
  a+b\,e^{-c\,\Delta E_\textrm{so}}\, \cos(d\,\Delta E_\textrm{so})
  .
\end{eqnarray}
This fitting function represents the ratio $S_\textrm{so}(\omega)/S_\textrm{no-so}(\omega)$ according to Eq.~\eqref{eq:hhg_so}, where an additional damping term $e^{-c\,\Delta E_\textrm{so}}$ has been introduced to account for the reduced tunnel ionization rate of $4p_{1/2}$ due to the increased ionization potential.

In the limit of large $\Delta E_\textrm{so}$, the orbital $4p_{1/2}$ is so tightly bound that the tunnel ionization rate drops basically to zero and, hence, it does not contribute anymore to the HHG spectrum.
As a consequence, krypton reduces effectively to a single-shell system with one active $4p_{3/2}$ shell.
This fact can be nicely seen in Fig.~\ref{f2} for large $\Delta E_\textrm{so}$, where the normalized yield drops independently of $\lambda$ to $0.49 \pm 0.02$ (extracted from the fitting parameters $a$ and $b$).
From Eq.~\eqref{eq:hhg_so}, which is based on a simple model, one would expect for large $\Delta E_\textrm{so}$ the ratio $5/9=0.5\bar{5}$, which is very close to the extracted value 0.49.
The discrepancy between these two values is probably due to the model assumption that only the $m^L=0$ part of the coupled orbitals contributes to tunnel ionization, and not the $m^L=\pm 1$ parts.

The frequency of the oscillations in Fig.~\ref{f2} is according to Eq.~\eqref{eq:hhg_so} $t_c$---the time the electron spends in the continuum.
From the fitting procedure, we obtain $t_c = 0.64\ T_\lambda$ for all three wavelengths, where $T_\lambda$ is the corresponding cycle period.
This time agrees with classical calculations predicting a trajectory time of $0.65\,T_\lambda$ for the trajectory with the maximum returning kinetic energy which is associated with generating the cut-off energy---the energy region we consider here.

Furthermore, the results from Fig.~\ref{f2} show that the spin-orbit splitting should not be too large when varying the wavelength, since the contrast in the oscillations decreases exponentially with $\Delta E_\textrm{so}$.
Too small $\Delta E_\textrm{so}$ are also not good, since they require a large wavelength range in order to cover with $t_c$ an entire spin-orbit period.



\subsection{Varying Driving Wavelength}
\label{s3.2}

An experimentally more realistic approach is to vary the driving wavelength $\lambda$.
The HHG yield changes now not just due to the spin-orbit factor in Eq.~\eqref{eq:hhg_so}.
Varying $\lambda$ also changes the cut-off energy and strongly affects the overall yield, which scales roughly with $\lambda^{-6.5}$~\cite{TaAu-PRL-2007,FoMa-PRL-2008,ShTr-PRL-2009} when keeping the pulse duration fixed with respect to the number of optical cycles.
Unfortunately, this wavelength scaling is much more pronounced than the modulation due to spin-orbit.
However, increasing wavelength results in a monotonically decreasing yield whereas the spin-orbit coupling (as seen in Sec.~\ref{s3.1}) can also lead to an increasing yield (for certain pulse parameters).

\begin{figure}[t!]
  \centering  
  \rmpdfinfo
  \includegraphics[width=\linewidth]{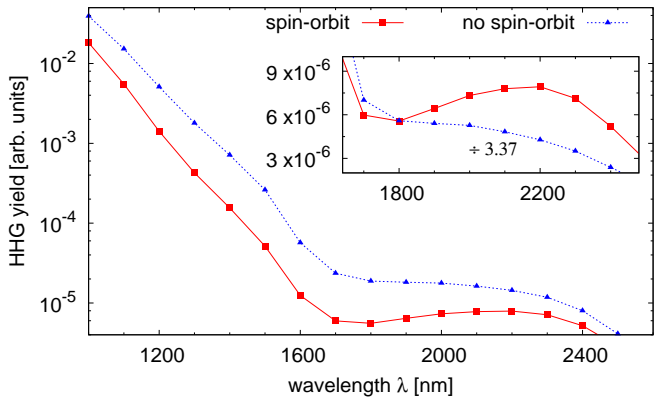}
  \caption{(color online) 
    The absolute HHG yield around the HHG cut-off energy as a function of the NIR wavelength with (solid-red line with squared dots) and without (dashed-blue line with triangular dots) spin-orbit coupling.
    The inset zooms into the region where the HHG yield increases.
    The non-spin-orbit curve is divided by 3.37 in the inset in order to put it on the same scale as the spin-orbit curve for better comparison of the different trends.
    The pulse parameters are the same as in Fig.~\ref{f2}.
  }
  \label{f3}
\end{figure}

Figure~\ref{f3} shows the absolute HHG yields around the HHG cut-off energy as a function of wavelength for krypton atoms with (solid-red line) and without (dashed-blue line) spin-orbit coupling ("with spin-orbit coupling" means $\Delta E_\textrm{so}=0.67$~eV).
Note that we keep the absolute pulse duration constant---not the pulse duration with respect to the wavelength.
Around $\lambda \sim 1700-2400$~nm we see in both scenarios, independent of the spin-orbit coupling, a plateau in the HHG yield~\footnote{
The reason for this reduced $\lambda$-dependence is the increasingly few-cycle character of the pulse. 
This also means there will be a CEP-dependence which is not the focus of this study where always the following pulse form is used: $\cos(\omega t)\ e^{-t^2/d^2}$.
This plateau is not expected for a multi-cycle pulse.}.
In this region, the wavelength dependence is relatively weak such that the spin-orbit dependence becomes very visible.
Without spin-orbit coupling the HHG yield slightly decreases whereas with spin-orbit coupling we actually see an increase in the yield by a factor $\sim 1.5$.
A weak wavelength dependence is advantageous for identifying spin-orbit effects which is relatively weak compared to the general $\lambda^{-5.5}$ to $\lambda^{-6.5}$ scaling.

The increase appears around this wavelength for krypton because for $\lambda=1600$~nm and larger the time of flight $t_c$ of the returning electron exceeds $\frac{T_\textrm{so}}{2}$ and the $4p_0$ character of the ionic hole at recombination increases again.
For $0 \leq t_c \leq \frac{T_\textrm{so}}{2}$ the $4p_0$ character first falls monotonically with $t_c$ before it increases again.
In Fig.~\ref{f2}(b) this effect is very well visible for the different wavelengths.
At $\lambda=1500$~nm the relative HHG yield is at a minimum for the experimental spin-orbit splitting, and at $\lambda=2000$~nm the relative yield is increasing again.
Generally, one finds for $n\,T_\textrm{so} \leq t_c \leq \frac{2n+1}{2}\,T_\textrm{so}$ the yield decreases and for $\frac{2n+1}{2}\,T_\textrm{so} \leq t_c \leq (n+1)\,T_\textrm{so}$ the yield increase with $n\in\mathbb{N}$.

\section{Conclusion}
\label{s4}

We have shown that spin-orbit coupling does affect the HHG yield due to multiorbital tunnel ionization of the outermost $p_{3/2}$ and $p_{1/2}$ states, which is followed by a non-stationary hole motion.
In contrast to molecular systems with multiorbital contributions, the angular dependence of the HHG emission is unaffected, since the total angular momentum state of the entire system is always $M^J=0$ and the electron spin does not change by a dipole transitions~\footnote{Dipole transitions where the spin changes can, in principle, occur but they are strongly suppressed compared to the non-spin-changing transitions.}.

Furthermore, we showed that by changing the strength of the spin-orbit coupling we can see clear oscillations in the HHG yield, which agrees well with the simple picture that the hole has initially a $p_0$ character but when the electron recombines this is only partly true leading to a reduced recombination probability.
Experimentally, the spin-orbit motion of the hole can be seen in HHG by changing the driving wavelength.
However, macroscopic aspects like phase-matching have to be understood, since they can lead to strong enhancements and reduction in the HHG yield~\cite{GaTa-JPB-2008_review}.
Large spin-orbit splittings, on the one side, lead to fast oscillations as a function of the wavelength but also suffer in visibility due to an exponentially decreasing strength in these oscillations.
Small spin-orbit splittings, on the other side, require a large range of driving wavelengths to cover a full spin-orbit period with the times the electron spends in the continuum before recombining.

Even though it seems to be a non-trivial challenge to show experimentally the influence of the spin-orbit effect, it does exist and it influences how the HHG yield scales with the wavelength.

\acknowledgments
This work has been supported by the Deutsche Forschungsgemeinschaft (DFG) under grant No. SFB 925/A5.

\appendix
\section{Influence of $3d$ and $4s$ orbitals}
\label{app1}
\begin{figure}[ht!]
  \centering  
  \rmpdfinfo
  \includegraphics[clip,width=\linewidth]{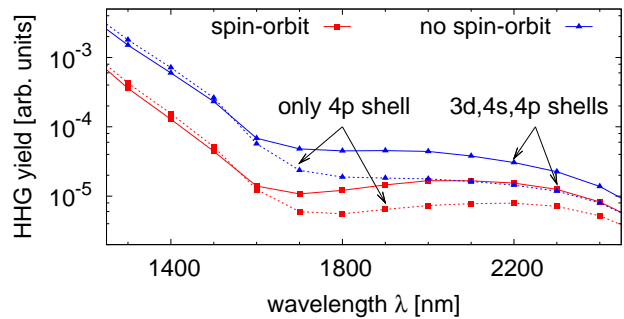}
  \caption{(color online)  
    The absolute HHG yield around the HHG cut-off energy as a function of wavelength with (red lines) and without (blue lines) spin-orbit splitting.
    In each case only the $4p$ shell (dotted lines; see Fig.~\ref{f2}) or the $3d, 4s$, and $4p$ shells (solid lines) are active.
  }
  \label{f4}
\end{figure}

The results shown in Sec.~\ref{s3} were all calculated by considering only the $4p$ shell and the spin-orbit coupling within this shell.
In Fig.~\ref{f4} we show the influence of the deeper lying $4s$ and $3d$ shells for the HHG yield.
We see that around $\lambda \sim 1600$~nm the HHG yield starts to differ and at $\lambda \sim 1800$~nm the enhancement when including the $4s$ and $3d$ shells has grown to a factor 2.5.
This enhancement is independent of the spin-orbit consideration and exists equally in the no-spin-orbit scenario.

Furthermore, the enhancement starts exactly at the wavelength where the HHG cut-off energy reaches the ionization potential of the $3d$ shell at 93.8~eV~\cite{xdb}.
To be precise, at $\lambda=1600$~nm (1800~nm) with a cycle-averaged peak intensity of around $8.8\times 10^{13}$~W/cm$^2$ ($E_\textrm{max}=0.05$~a.u.), the HHG cut-off energy is 80~eV (98~eV).
However, since the enhancement happens equally with and without spin-orbit coupling the overall conclusions are unchanged.

\bibliographystyle{apsrev4-1}
\bibliography{amo,books,notes,coldgases}


\end{document}